\documentclass[]{tPHM2e}

\newcommand{\codo}{Ca$_{10}$(Fe$_{1-x}$Co$_x$As)$_{10}$(Pt$_3$As$_8$)}
\newcommand{\nido}{Ca$_{10}$(Fe$_{1-x}$Ni$_x$As)$_{10}$(Pt$_3$As$_8$)}
\newcommand{\cudo}{Ca$_{10}$(Fe$_{1-x}$Cu$_x$As)$_{10}$(Pt$_3$As$_8$)}
\newcommand{\ptdo}{Ca$_{10}$(Fe$_{1-x}$Pt$_x$As)$_{10}$(Pt$_3$As$_8$)}
\newcommand{\parent}{Ca$_{10}$(FeAs)$_{10}$(Pt$_3$As$_8$)}
\newcommand{\zva}{Ca$_{10}$(FeAs)$_{10}$(Pt$_4$As$_8$)}

\newcommand{\mdo}{Ca$_{10}$(Fe$_{1-x}M_x$As)$_{10}$(Pt$_3$As$_8$)}
\newcommand{\redo}{(Ca$_{1-x}RE_x$)$_{10}$(FeAs)$_{10}$(Pt$_3$As$_8$)}

\begin{document}



\markboth{St\"urzer, Kessler, and Johrendt}{\mdo}

\articletype{}

\title{Superconductivity by transition metal doping in Ca$_{10}$(Fe$_{1-x}M_x$As)$_{10}$(Pt$_3$As$_8$) ($M$ = Co, Ni, Cu)}

\author{Tobias St\"urzer, Fabian Kessler, and Dirk Johrendt$^\ast$ \thanks{$^\ast$Corresponding author. Email: johrendt@lmu.de\vspace{6pt}}\\\vspace{6pt}  
{\em{Department Chemie, Ludwig-Maximilians-Universit\"at M\"unchen, D-81377 M\"unchen, Germany}}}

\maketitle

\begin{abstract}
We report the successful substitution of cobalt, nickel, and copper for iron in the 1038 phase parent compound \parent\ yielding \codo, \nido, and \cudo, respectively. Superconductivity is induced in Co and Ni doped compounds reaching critical temperatures up to 15 K, similar to known Pt substituted \ptdo, whereas no superconductivity was detected in \cudo. The obtained $T_c(x)$ phase diagrams are very similar to those of other iron arsenide superconductors indicating rather universal behaviour despite the more complex structures of the 1038-type compounds, where the physics is primarily determined by the FeAs layer.      

\bigskip


\end{abstract}

\section*{Introduction}

\begin{figure}[h]
\centering
\includegraphics[width=9cm]{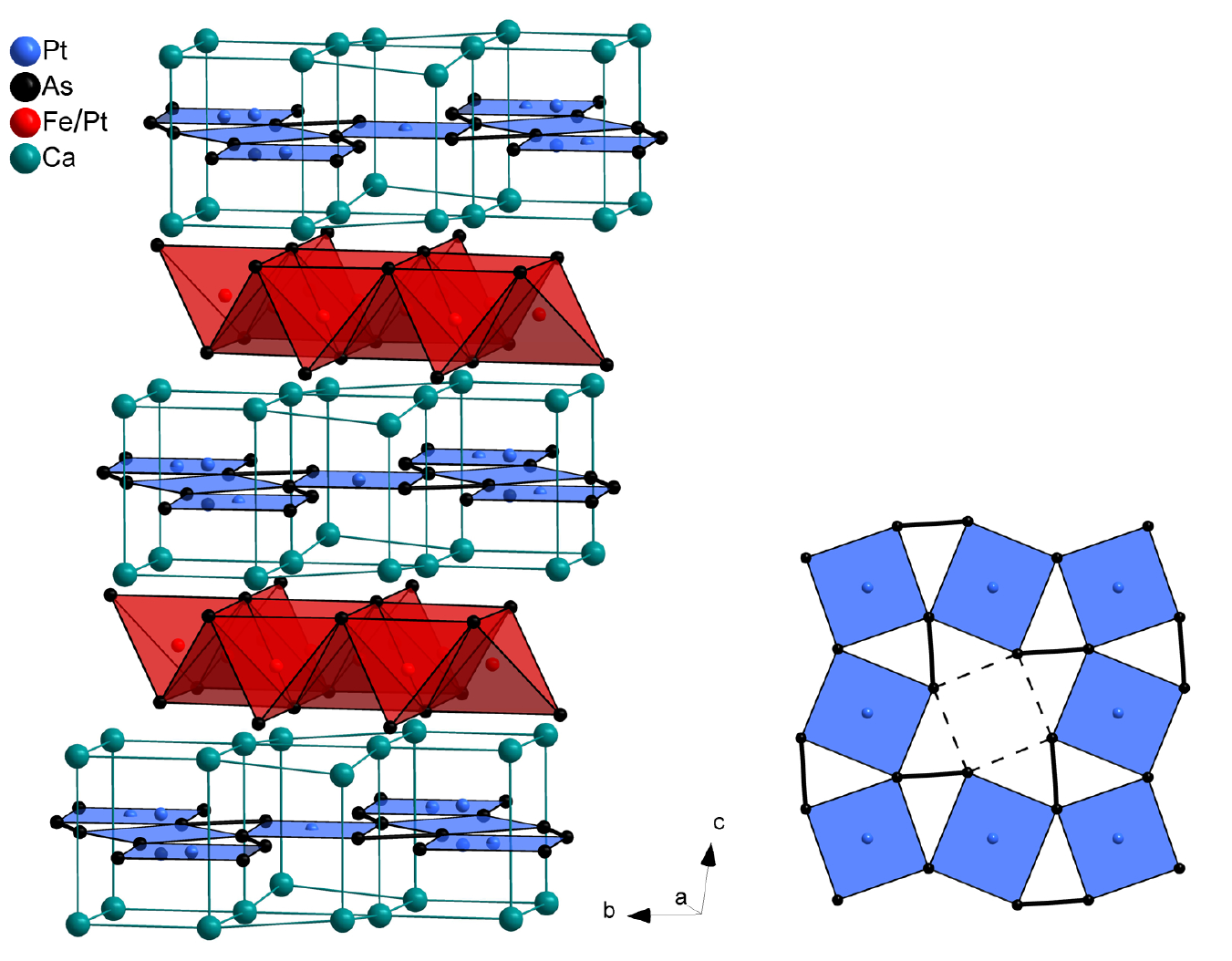}
\caption{\label{fig:structure} Crystal structure of the 1038 parent compound  \parent.}
\end{figure}

Superconductivity in iron arsenides emerges from antiferromagnetic metallic parent compounds in the course of suppressing the magnetic ordering by chemical doping or pressure \cite{Kamihara-2008, Rotter-2008, Alireza-2009, Johnston-2010, Johrendt-2011, Stewart-2011}, resulting critical temperatures ($T_c$) up to  55 K in Sm(O$_{1-x}$F$_x$)FeAs \cite{Ren-2008}. Relationships between the magneto-structural phase transition and superconductivity in iron arsenides have intensively been studied \cite{Cruz-2008,Zhao-2008,Rotter-2008-2, Li-2009,Mcguire-2008, Paglione-2010,Johrendt-2011,Fernandes-2014}. In 2011 the new superconductors \ptdo\ (1038 phase, space group $P\overline{1}$) and polymorphic \zva\ (1048-phases, space groups $P4/n$, $P2_1/n$, $P\overline{1}$) with critical temperatures up to 35 K were discovered \cite{Loehnert-2011,Ni-2011,Kakiya-2011}. This new class recently expanded by analogous compounds with iridium (Ir1048) \cite{Kudo-2013} and palladium (Pd1038) \cite{Hieke-2013} instead of platinum. Due to the presence of the second metal pnictide layer Pt$_z$As$_8$ ($z$ = 3, 4) next to FeAs, as well as the low symmetry of these compounds (space group $P\overline{1}$) they were initially considered as rather peculiar representatives of the iron arsenide family. However, recent low temperature X-ray structural data together with $\mu$SR spectra \cite{Stuerzer-2013} as well as neutron diffraction \cite{Sapkota-2014} revealed a lattice distortion and magnetic phase transition at 120 K , proving that \parent\ is the parent compound of this branch of the iron arsenide family. Thereby closely related superconductors like \ptdo\ ($T_c^{\rm max}$ = 14 K), \zva\ ($T_c^{\rm max}$ = 35 K) and \redo\ ($T_c^{\rm max}$ = 35 K) can be derived from this common parent by direct, indirect, and electronic doping, respectively \cite{Stuerzer-2012,Stuerzer-2014,Ni-2013}.  Fig.~\ref{fig:structure} depicts the structure of the 1038 parent compound  as well as a section of the Pt$_3$As$_8$ layer.

In this paper we report the crystal structures and superconductivity of transition metal doped \mdo\ with $M$ = Co, Ni, and Cu with critical temperatures up to 15 K. Our results clearly establish that the 1038/1048 superconductors are more than an exotic exception, but another, even structurally more complex, branch of the iron arsenide family.


\section*{Experimental} 

Polycrystalline samples of Co-, Ni-, and Cu-doped calcium platinum iron-arsenides were synthesized from the elements as described in \cite{Loehnert-2011}, and characterized by X-ray powder diffraction using the Rietveld method with TOPAS \cite{Topas}. Compositions were determined within errors of 10\% by X-ray spectroscopy (EDX). Superconducting properties were determined using an $ac$-susceptometer at 1333 Hz in the temperature range of 3.5 K to 300 K at 3 Oe. Magnetic measurements were additionally performed on a Quantum Design MPMS XL5 SQUID magnetometer which allowed for measurements with fields up to 50 kOe at temperatures between 1.8 K and 300 K. Temperature dependent resistivity measurements between 3.5 K and 300 K were carried out using a standard  four-probe method.\bigskip

\section*{Results}

\begin{figure}[h]
\begin{center}
\subfigure[]{\includegraphics[width=0.49\textwidth]{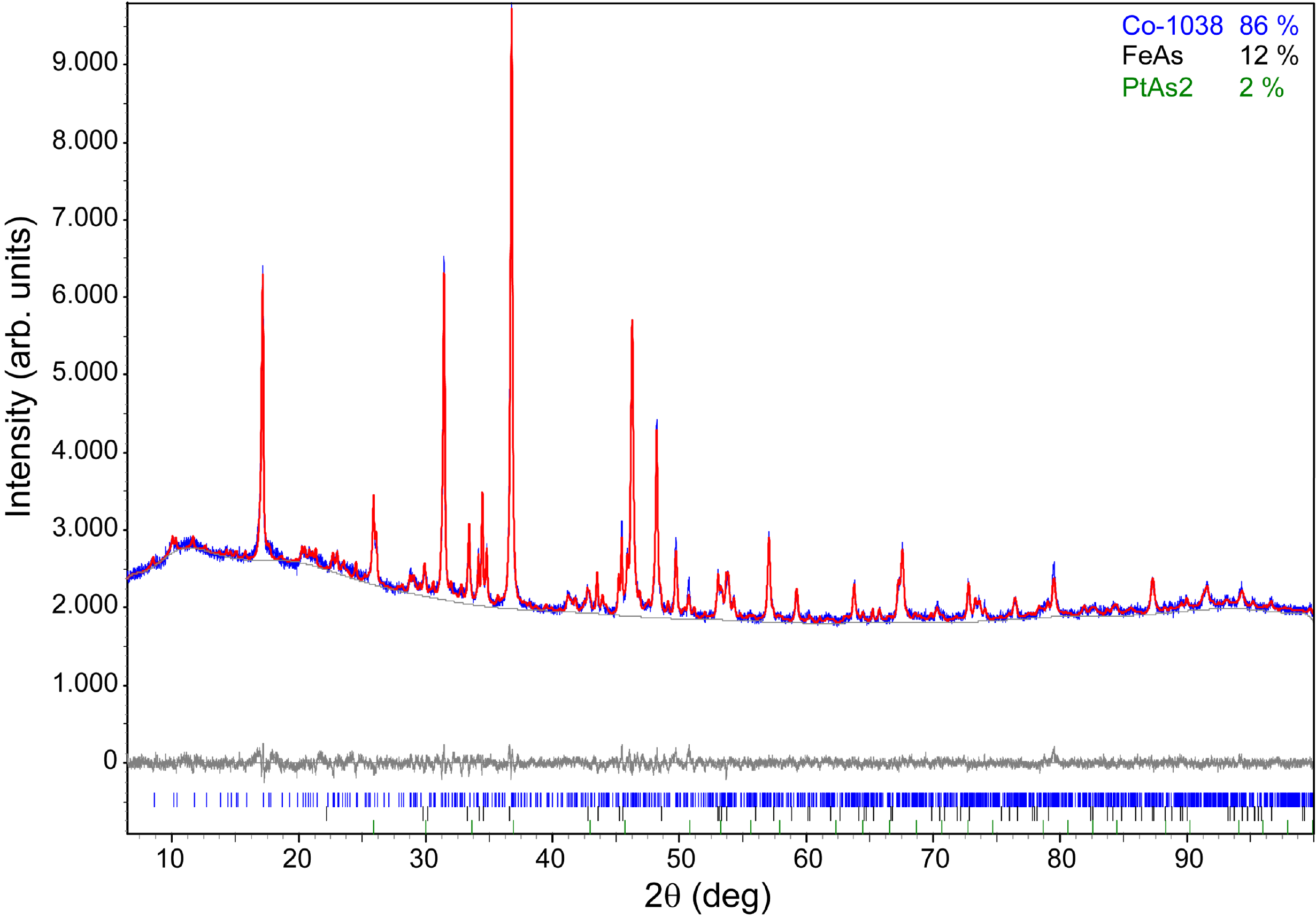}}
\subfigure[]{\includegraphics[width=0.49\textwidth]{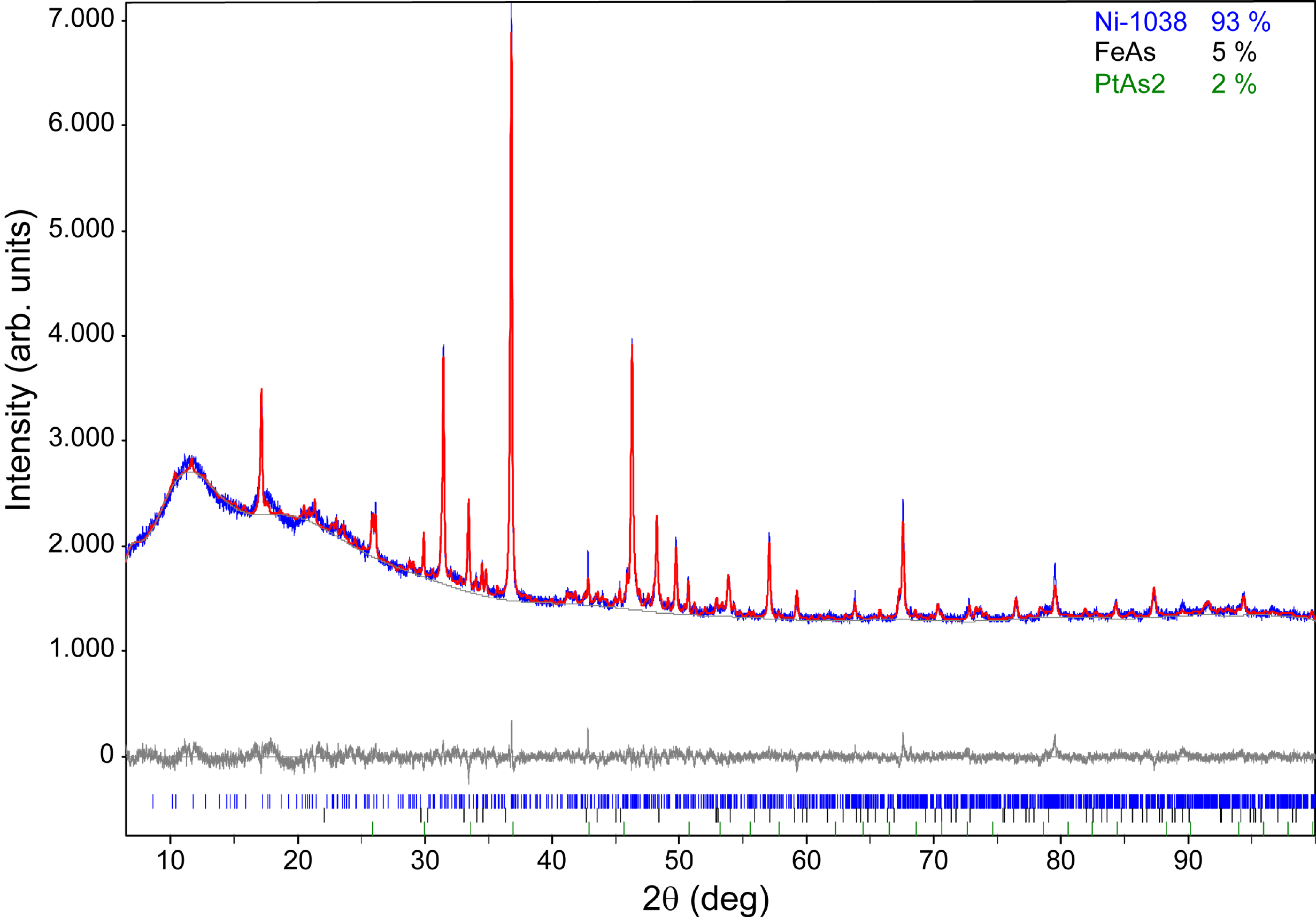}}
\caption{\label{fig:rietveld} X-ray powder patterns (blue lines) with Rietveld fits (red lines) of \mdo\ with $M$ = Co (a) and Ni (b).}
\end{center}
\end{figure}

Fig.~\ref{fig:rietveld} shows X-ray powder patterns with Rietveld fits of \mdo\ samples with $M$ = Co, Ni and $x$ = 0.1. The diffraction patterns are completely described with the 1038-phase structure model \cite{Loehnert-2011} and minor amounts of impurity phases FeAs and PtAs$_2$. The high sample quality implies smooth incorporation of cobalt, nickel, and copper into the 1038 structure. The amount of impurity phases increase at higher doping levels, which indicates solubility limits of $x \approx $ 0.25 for Co and Ni, as well as $x \approx$ 0.12 for Cu, respectively, in the 1038-type structure. Thus, the fully substituted compounds Ca$_{10}$(CoAs)$_{10}$(Pt$_3$As$_8$), Ca$_{10}$(NiAs)$_{10}$(Pt$_3$As$_8$), Ca$_{10}$(CuAs)$_{10}$(Pt$_3$As$_8$) and Ca$_{10}$(FeAs)$_{10}$(Ni$_3$As$_8$) were not accessible by solid state synthesis, although the Pd analogue Ca$_{10}$(FeAs)$_{10}$(Pd$_3$As$_8$) was recently reported \cite{Hieke-2013}.

\begin{figure}[h]
\begin{center}
\subfigure[]{\includegraphics[width=0.49\textwidth]{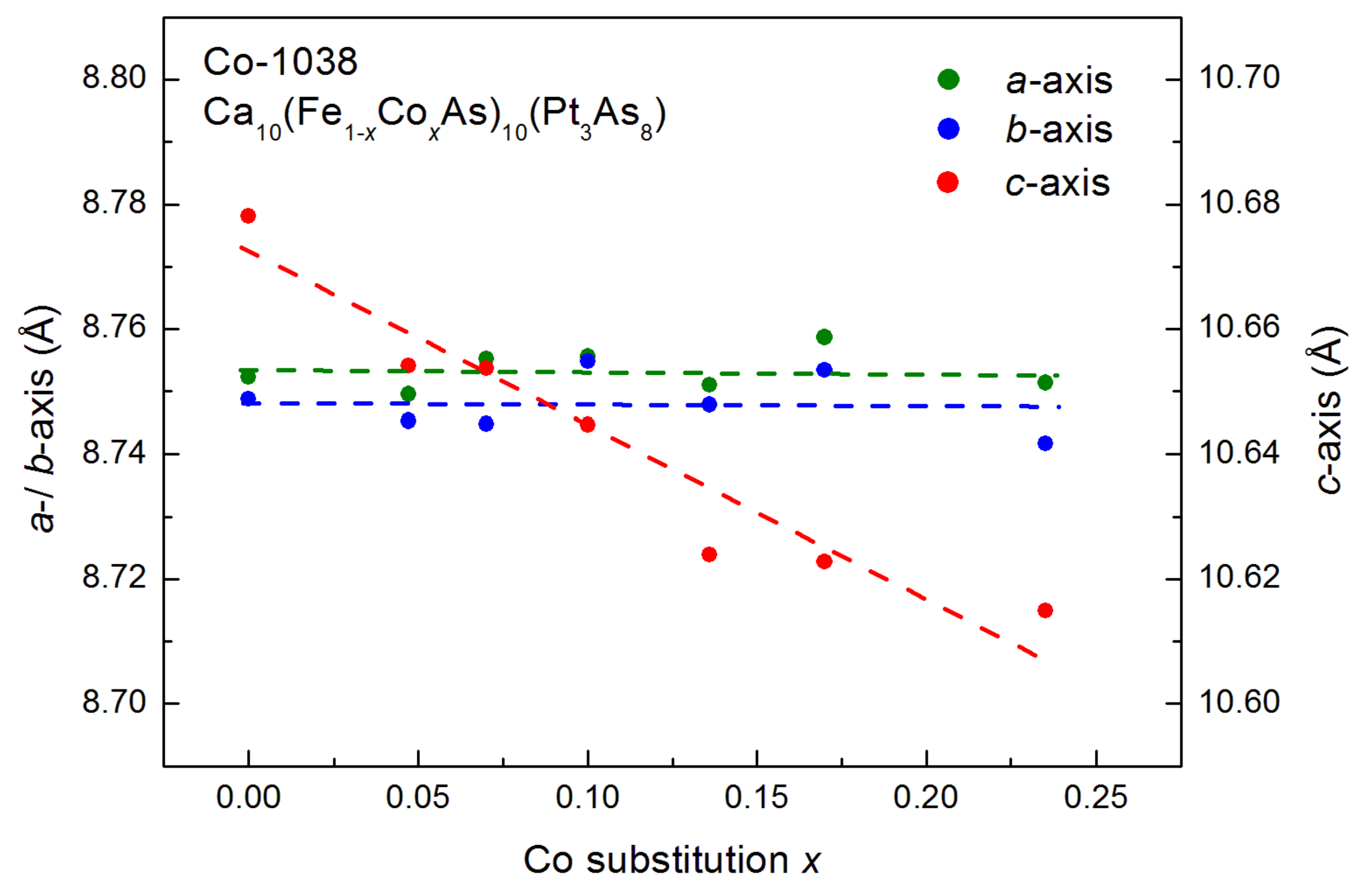}}
\subfigure[]{\includegraphics[width=0.49\textwidth]{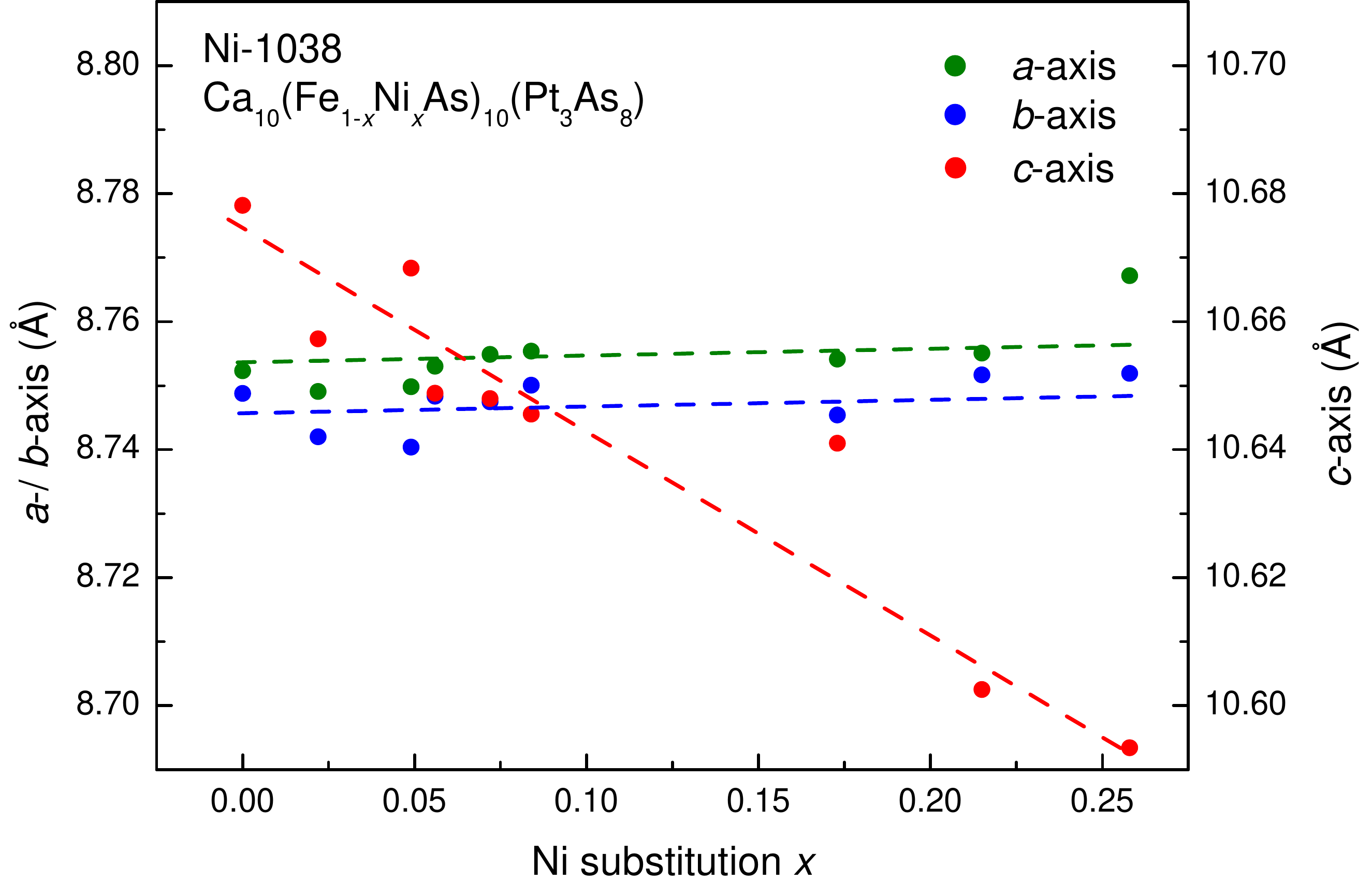}}
\caption{\label{fig:lattice} Lattice parameters of \mdo\ with $M$ = Co (a) and Ni (b) obtained from Rietveld fits.}
\end{center}
\end{figure}

Fig.~\ref{fig:lattice} illustrates the dependency of the lattice parameters on Co- or Ni-substitution from X-ray powder data refinement, revealing similar behaviour of both Co- and Ni-1038 to the 122-type compounds Ba(Fe$_{1-x}$Co$_x$)$_2$As$_2$ \cite{Sefat-2008} and Ba(Fe$_{1-x}$Ni$_x$)$_2$As$_2$ \cite{Li-2009}. The in-plane axes $a$ and $b$, as well as the unit cell angles (not shown) remain nearly constant within experimental accuracy, whereas $c$ decreases by about 0.8\% in the range $0 \leq x \leq 0.25$, accompanied with shrinking of the cell volume by approximately 0.8\%. At this point it should be noted, that, even if small, the analogous effects of Co- and Ni-doping to the 1038-structure may be indicative for a similar 3$d$ electron count localized at Co and Ni when doped to FeAs layers.\bigskip

An interesting aspect arises from the chemical similarity of platinum and nickel. Although the targeted position of Co when doped into the 1038 structure is the FeAs layer, the situation is more complicated in the case of Ni. In this system Fe/Ni mixing is expected, but Ni substitution to the Pt sites is as well imaginable as Ni occupancy at the Pt vacancies in the Pt$_3\square$A$_8$ layer. However, modifications of the Pt$_3\square$A$_8$ layers by additional Ni incorporation were supposed to produce 1048-type impurity phases, or at least changes of the $a$,$b$ lattice parameters. Susceptibility measurements would be a sensitive probe to detect even minimal traces of the 1048 phase due to its high $T_c$  well above 30 K. However, none of these phenomena have been observed. Moreover, additional EDX measurements gave no indication for Pt/Ni mixing. All these results clearly indicate that no modifications of the Pt$_3$As$_8$ layers occur, tantamount with Co-, Ni- or Cu-doping taking place only in the FeAs layers.

\begin{figure}[h]
\begin{center}
\subfigure[]{\includegraphics[width=0.49\textwidth]{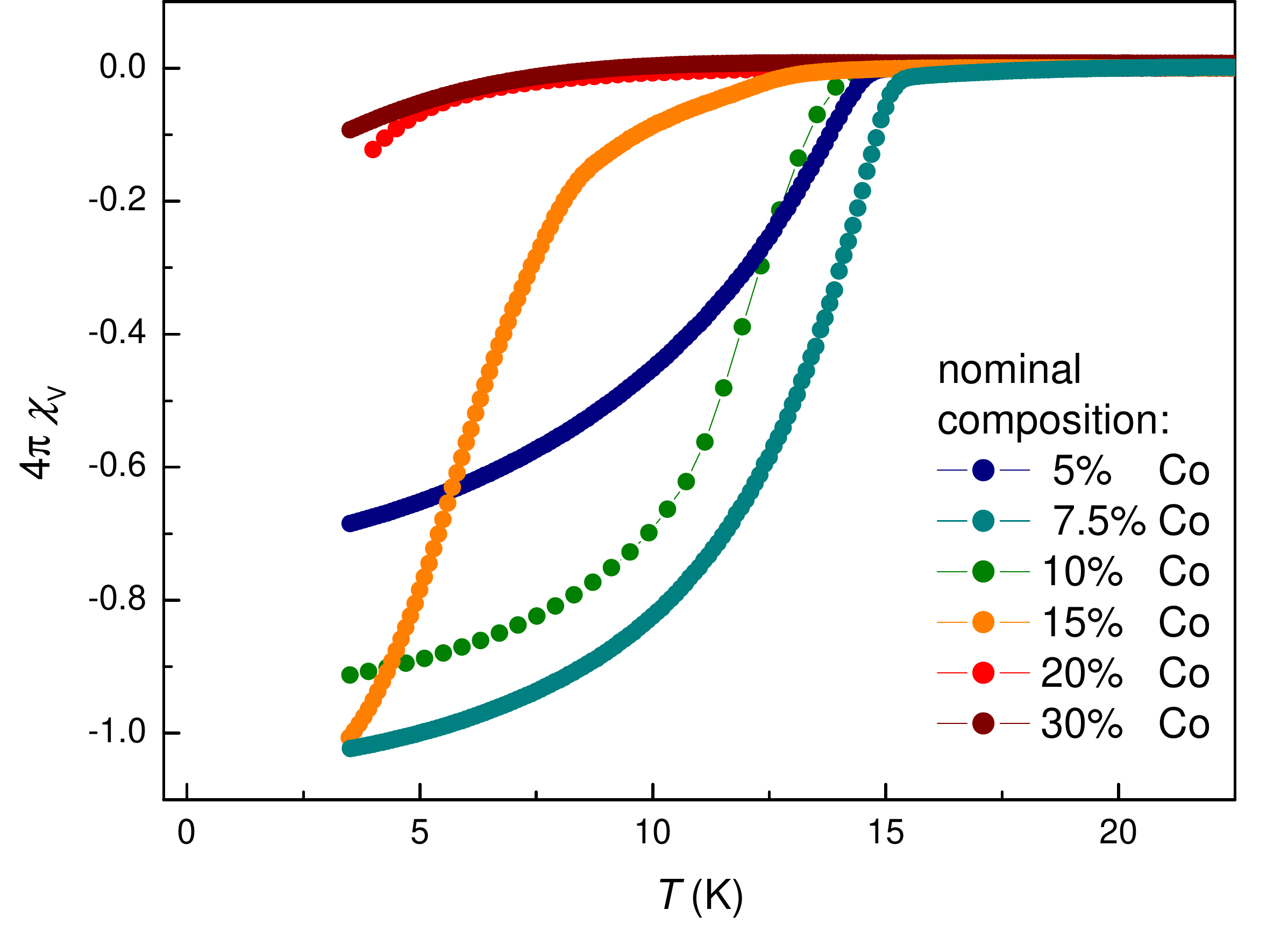}}
\subfigure[]{\includegraphics[width=0.49\textwidth]{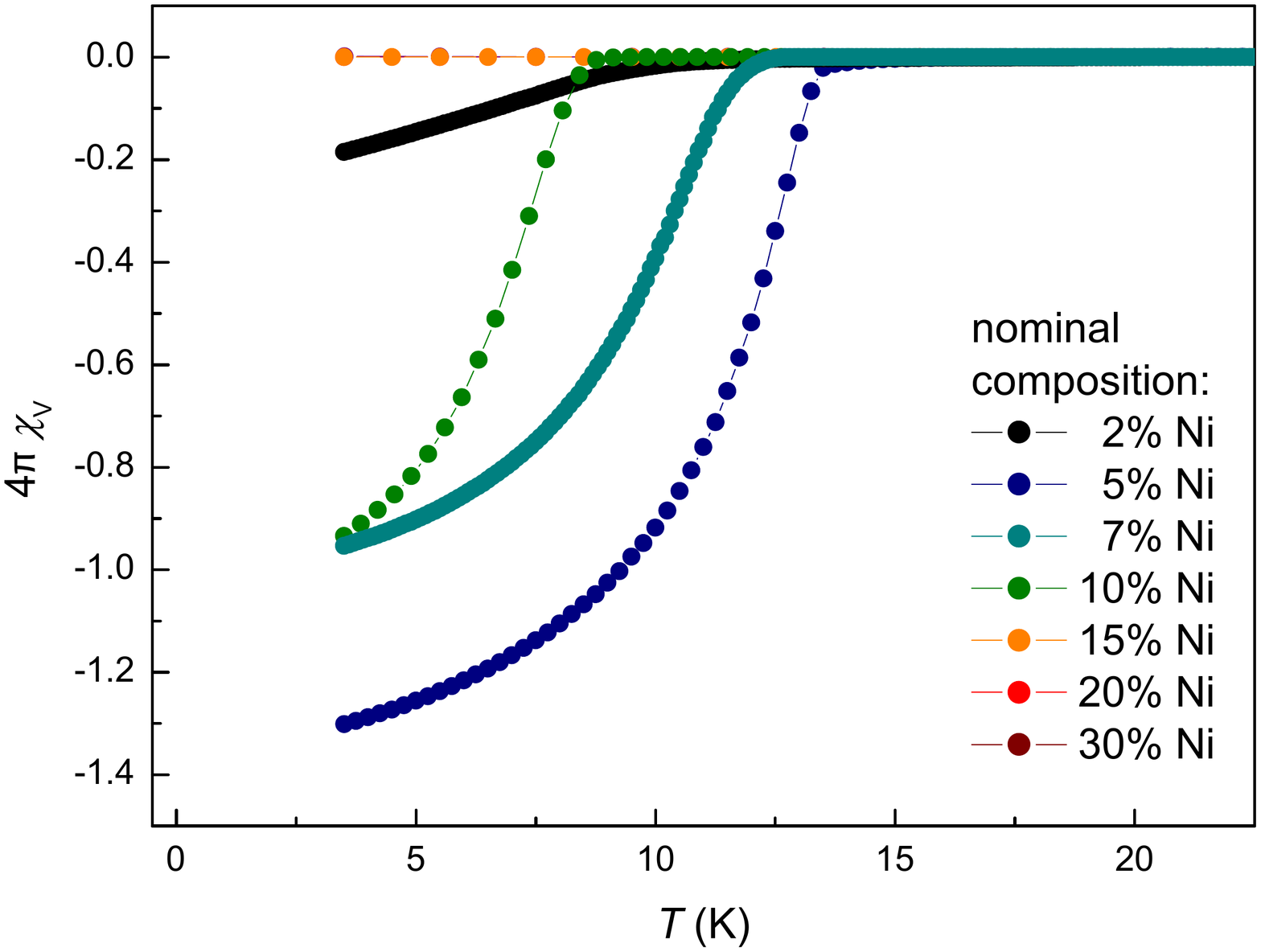}}
\caption{\label{fig:sus} $ac$-susceptibilities of \codo\ (a) and \nido\ (b).}
\end{center}
\end{figure}

Fig.~\ref{fig:sus} shows $ac$-susceptibility data of \mdo\  ($M$ = Co, Ni). Critical temperatures of Co-1038 reach 15.3 K at the optimal Co concentration $x$ = 0.075, whereas Ni-1038 maximum $T_c$ settles at 13.4 K for a doping level of $x$ = 0.05. Superconducting volume fractions indicate bulk superconductivity. In contrast to this finding no superconductivity was detected in \cudo, except of traces at the lowest doping level sythesized of $x$ = 0.02.\bigskip 

\begin{figure}[h]
\begin{center}
\subfigure[]{\includegraphics[width=0.49\textwidth]{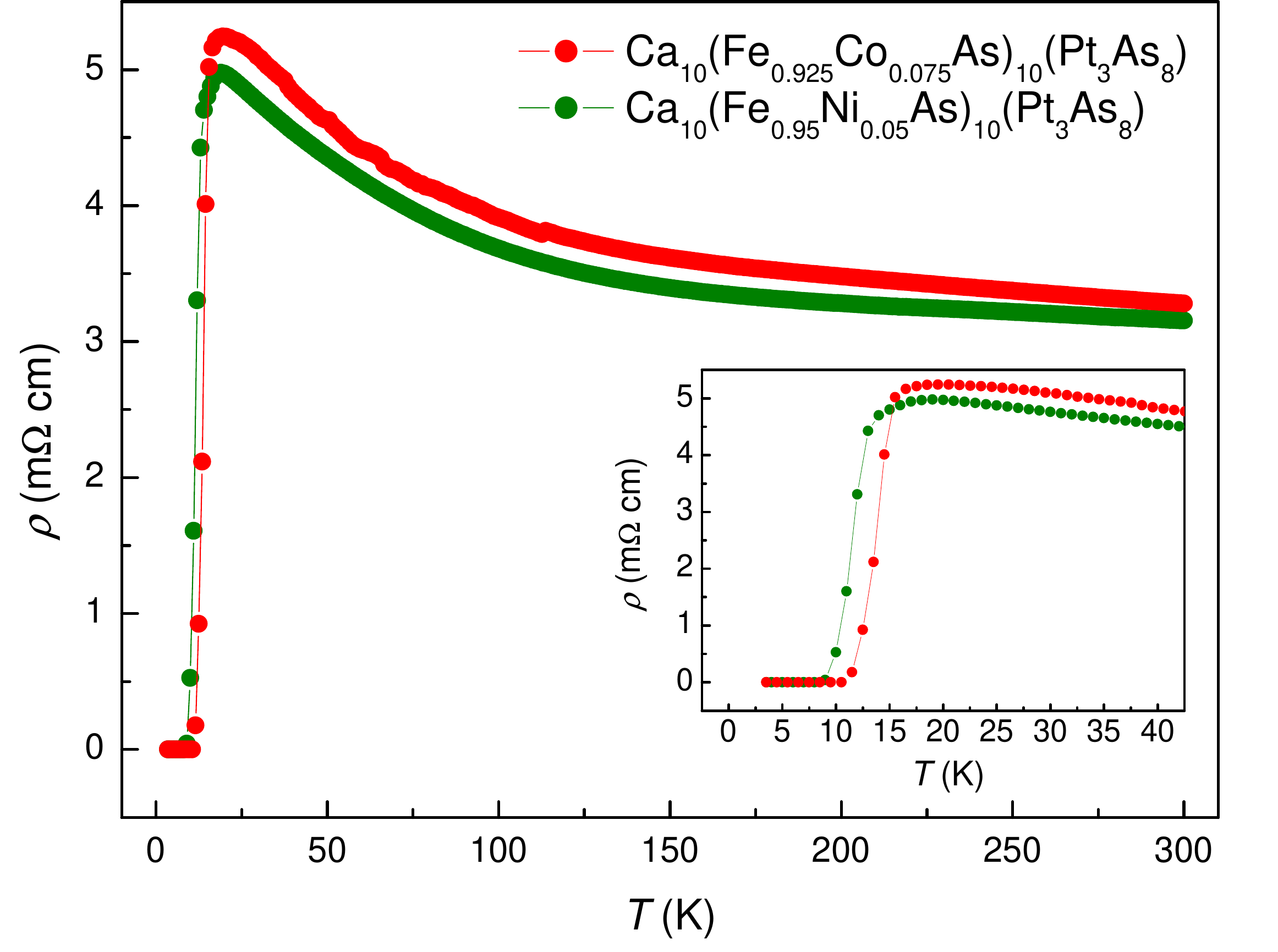}}
\subfigure[]{\includegraphics[width=0.49\textwidth]{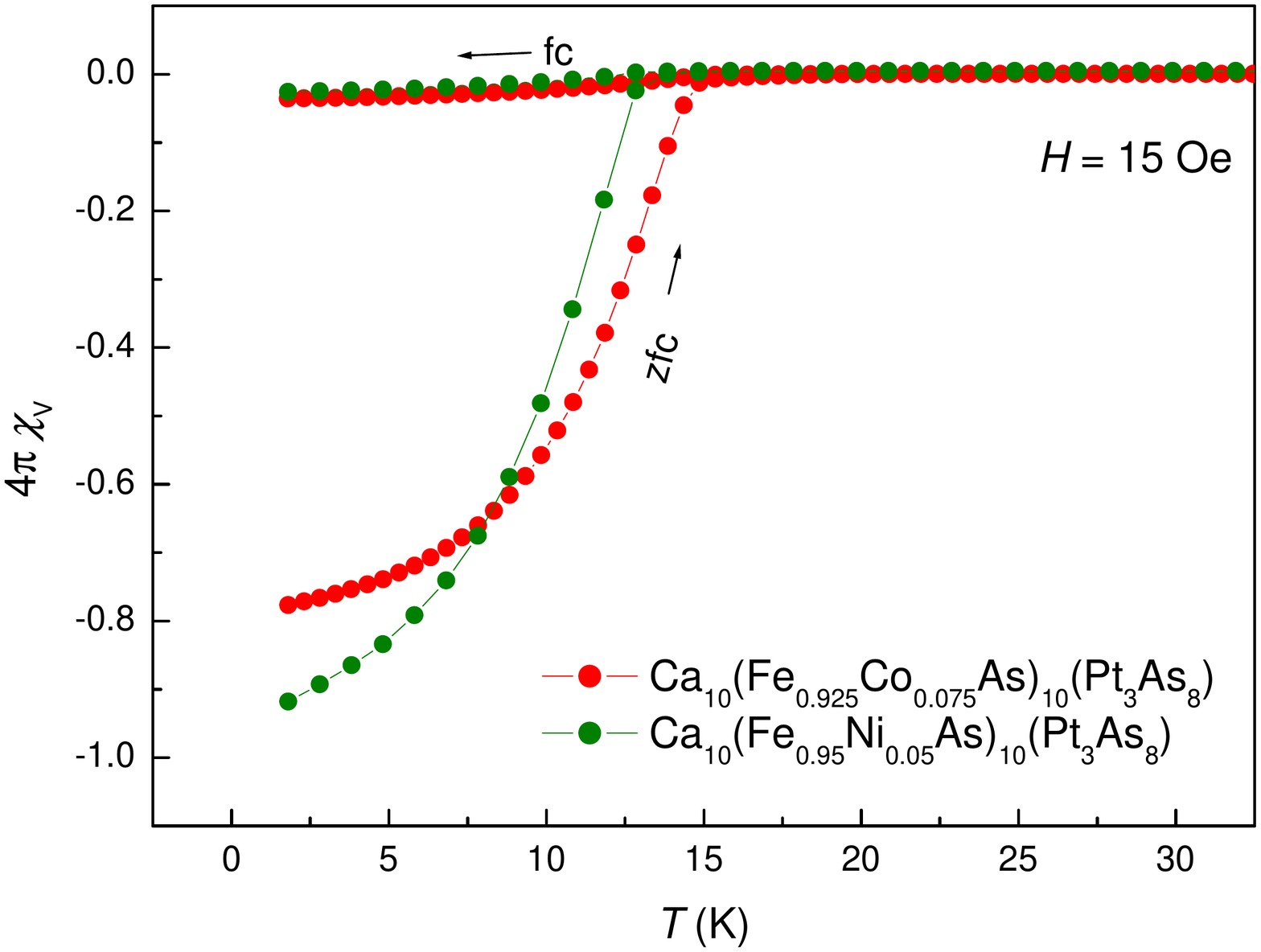}}
\caption{\label{fig:res} $dc$-resistivities (a) and low-field  $dc$-susceptibilities (b)  of \codo\ and \nido.}
\end{center}
\end{figure}

Fig.~\ref{fig:res} shows $dc$-electrical resistivity and low field $dc$-susceptibility data. The left panel (Fig.~\ref{fig:res}a) displays resistivity measurements performed with the maximum $T_c$ samples Ca$_{10}$(Fe$_{0.925}$Co$_{0.075}$As)$_{10}$(Pt$_3$As$_8$) and Ca$_{10}$(Fe$_{0.95}$Ni$_{0.05}$As)$_{10}$(Pt$_3$As$_8$). A steep drop to zero resistance is observed for both samples which coincide with critical temperatures from susceptibility measurements, respectively. The absolute values of the specific resistivity are in the range of poor metals as typical for iron arsenides in the normal state. Remarkably, in our case the normal state resistivity increases with decreasing temperature resembling a semiconductor like behaviour. This temperature dependence of the specific resistivity is different from 1111- and 122-type iron arsenide superconductors, but was also found for the parent compound \parent. Similar results have also been reported for \ptdo\ \cite{Xiang-2012}. The right panel (Fig.~\ref{fig:res}b) shows the low-field $dc$-susceptibility of Co- and Ni-1038 samples with the highest $T_c$. Almost 100\% shielding at low temperatures proves bulk superconductivity, while the Meissner-signal is rather small.  

\begin{figure}[h]
\begin{center}
\includegraphics[width=9cm]{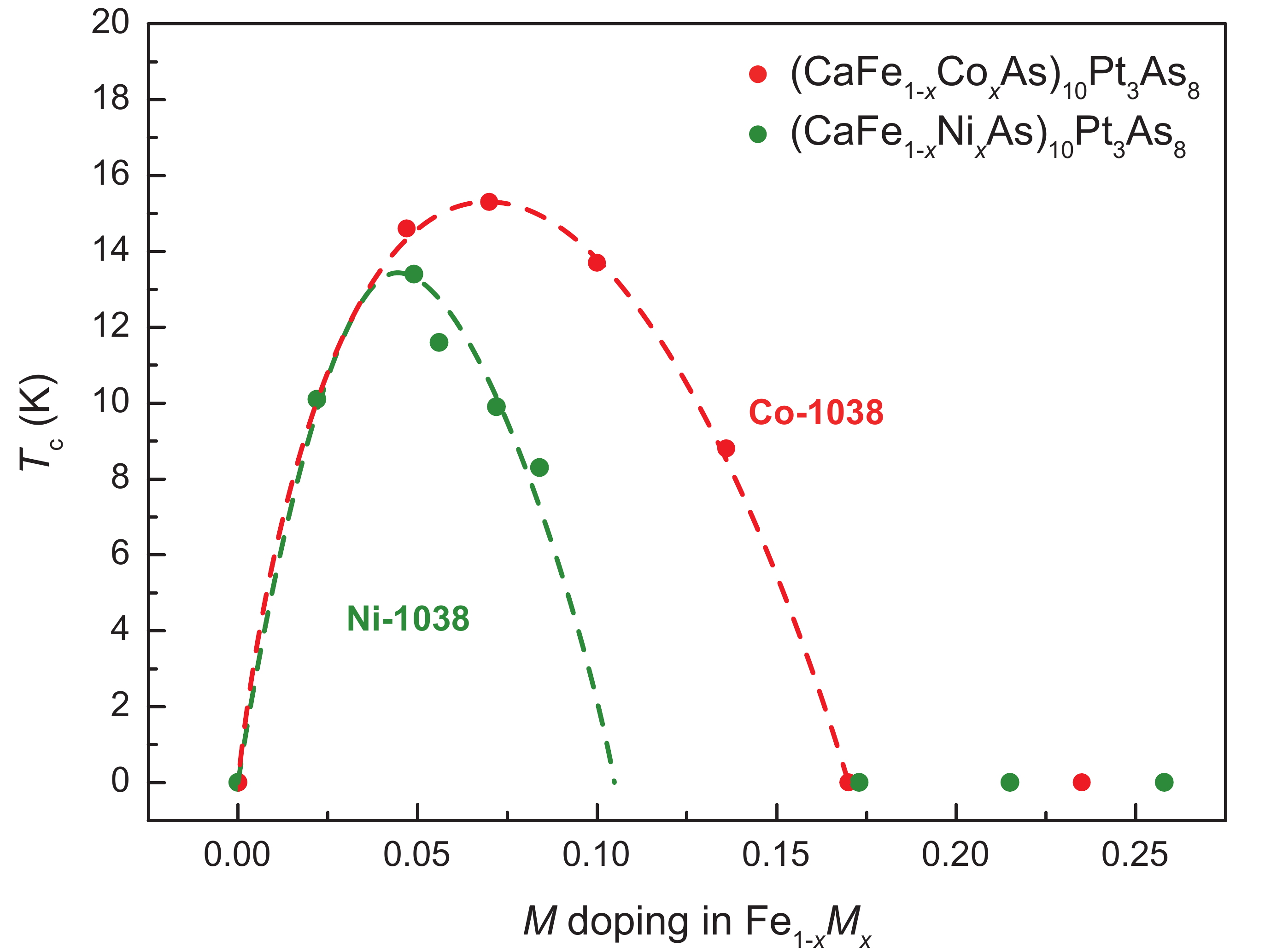}
\caption{\label{fig:phasedia} $T_c(x)$ phase diagrams of \codo\ and \nido.}
\end{center}
\end{figure}

Doping dependent critical temperatures for both Co- and Ni-1038 are compiled in Fig.~\ref{fig:phasedia}. Superconductivity is induced by small transition metal doping to the iron sites. \codo\ reveals a dome like $T_c(x)$ dependency, reaching a maximum of 15.3 K for the optimal doping level $x$ = 0.075. The critical temperature distinctly drops if $x$ exceeds 0.075. However, full suppression of superconductivity upon high Co concentrations could not be archived within the solubility limit of Co. Samples with a nominal composition of $x$ = 0.2 and $x$ = 0.3 show only traces of superconductivity which may come from inhomogeneously distributed cobalt. Therewith the superconducting dome of Co-1038 is remarkably similar to the Co-doped 122-compounds Ba(Fe$_{1-x}$Co$_x$)$_2$As$_2$ \cite{Sefat-2008,Canfield-2010}.
 
The Ni-1038 compounds reveal narrower dependency of the critical temperatures from the substitution level, featuring a maximum $T_c$ of 13.4 K at $x$ = 0.05. This finding is in line with the additional electron of Ni with respect to Co, giving rise of an increased electron doping at same transition metal substitution \cite{Canfield-2010}.  $T_c$ rapidly decreases at higher Ni concentrations until superconductivity is completely suppressed at 17 \% Ni. Thus, \nido\ reveals very similar properties than its homologue \ptdo. In this context Ni-1038 is supposed to be a more suitable system to study its properties in the overdoped regime due to the higher solubility of Ni. \bigskip

\section*{Conclusion}

In conclusion we have demonstrated bulk superconductivity in the 1038 compounds \mdo\ doped by $M$ = Co or Ni, with critical temperatures up to 15.3 K in Co-1038 ($x$ = 0.075) and 13.4 K in Ni-1038 ($x$ = 0.05), respectively. Superconducting properties in both compounds were evidenced by $ac$- and $dc$-susceptibility as well as $dc$-resistivity data. Moreover no superconductivity was evident in Cu-doped samples \cudo. The dependency of $T_c$ of the substitution level $x$ reveals similar behaviour than in known directly doped 122-type iron arsenides Ba(Fe$_{1-x}M_x$)$_2$As$_2$ with $M$ = Co, Ni, Cu, whereas the comparatively narrow superconducting dome of Ni-1038 is indicative for an increased electron doping contribution of Ni with respect to Co-1038. Our results clearly show the close resemblance of calcium platinum iron arsenides to other iron arsenide compounds, giving evidence that established doping methods to induce superconductivity are abundantly applicable also to more complex systems like the 1038 and 1048 materials. 

\section*{Acknowledgement}

This work was financially supported by the German Research Foundation (DFG) within the priority program SPP1458. 

\bibliographystyle{tPHM}


\end{document}